\documentclass[12pt]{article}
\usepackage[german,english]{babel}   
\usepackage{graphics}
\usepackage{graphicx}
\newcommand{\et}{{\em et al}}

\newcommand{\rms}{$rms$-radius}
\newcommand{\rmss}{$rms$-radii}

\newcommand{\qrt}{$q^2 \langle r^2 \rangle /6$}
\newcommand{\rt}{$\langle r^2 \rangle$}
\newcommand{\rf}{$\langle r^4 \rangle$}

\newcommand{\cd}{Coulomb distortion}

\begin{document}

\vspace*{0.2cm}
\begin{center}
{\bf \LARGE On the rms-radius of the proton}\\
\vspace{1cm} 
{\bf \large
Ingo  Sick}\\
\vspace{0.5cm}
{\small \sl 
Dept. f\"ur Physik und Astronomie,
Universit\"at Basel  \\ CH-4056 Basel, Switzerland \\}
\end{center}
\vspace{1.2cm}                 
\normalsize
\begin{center} Abstract \end{center}
\small
\begin{minipage}[t]{1.0cm}
\hfill
\end{minipage}
\begin{minipage}[t]{12cm}
We study the world data on elastic electron-proton 
scattering in order to determine the proton charge \rms. After accounting for the 
\cd\ and using a parameterization that allows to deal properly with the higher
moments we find a radius of $0.895 \pm 0.018 fm$, which is significantly larger
 than the radii used in the past.   
\end{minipage} \\
 \normalsize
\newpage
{\bf Introduction.} \hspace{0.3cm} 
The  root-mean-square ({\em rms}) radius  of the proton is a quantity of great 
interest for an understanding of
the proton; it describes the most important integral property concerning 
 its size. 
Accurate knowledge of the \rms~ of the charge distribution 
is needed for the interpretation of
high-precision measurements of  transitions in hydrogen atoms, studied in
connection with  measurements of fundamental constants \cite{Karshenboim01}; 
these measurements
recently have made great progress, and are now limited by the accuracy with
which the proton radius is known \cite{Udem97}.  The  radius  is
also needed for the planned measurements of muonic X-ray transitions 
\cite{Kottmann01};
 these experiments can only scan a narrow frequency range, which must be chosen
according to the best value of the \rms~ presently known. 
  
  The proton \rms~ in the past in general has been determined  from elastic
  electron-proton scattering. The usual approach has been to employ the most
  accurate cross sections at low momentum transfer $q$, perform an experimental 
  separation of longitudinal (L, charge) and transverse (T, magnetic) 
  contributions. The resulting charge data as a function
  of $q^2$ are then fit with an appropriate function to get 
  the \rms~, {\em i.e.} the $q^2=0$ slope of the form factor \footnote{This
  quantity can be determined without making use of the nonrelativistic notion of
  the charge density as a Fourier transform of the form factor.}. 
  
  Alternative approaches have included theory-motivated fits such as given by
  the Vector Dominance Model (VDM) in combination with  dispersion relations.
  
  {\bf Past results.} ~~ 
  The initial electron scattering experiments on the proton were performed some
  40 years ago by the Hofstadter group at Stanford 
  \cite{Bumiller61,Janssens66}.
  \nocite{Borkowski74,Borkowski75,Simon80,Simon81} 
  \nocite{Albrecht66,Bartel66,Frerejacque66,Albrecht67,Bartel67,Bartel73}
  \nocite{Ganichot72,Kirk73,Murphy74,Berger71,Bartel70} 
   This data, mainly
  at medium $q$ and not low $q$, was fitted using  multi-pole form factors. 
  From the parameters of the fit  an \rms~ could be calculated. The resulting
  value of 0.81$fm$, which is still quoted in the literature, should have long
  been superseded by values coming from more precise 
  data at {\em lower} $q$ which are indeed sensitive to the \rms .
  
  In the seventies, accurate low-$q$ data, mainly measured at the Mainz 
  electron accelerator, became available 
  \cite{Borkowski74}-\cite{Simon81}. 
   After an L/T-separation, the data
  were usually fitted with a polynomial expansion of the form factor
  \begin{equation}
  G_e(q)=1 - q^2 \langle r^2 \rangle /6 + q^4 \langle r^4 \rangle /120 -...
  \end{equation}
   and, in general, a floating
  normalization of the individual data sets in order to produce the lowest
  $\chi^2$. The most prominent result was
  probably the one obtained by Simon {\em et al.} \cite{Simon80},
   $r_{rms} = 0.862 \pm 0.012~fm$.
  
  Occasionally, fits  with 2- or 4-pole expressions \cite{Borkowski75b} 
  were performed,
  and  significantly bigger values, {\em i.e.} $0.88 \pm 0.02 fm$ and $0.92 
  \pm 0.02  ~fm$ were found as compared to values determined at very low $q$
  \cite{Murphy74}. 
  The difference was partly understood \cite{Borkowski75b} as a consequence of
  different treatments of the \rf~ term. 
  
  In parallel, fits based on dispersion relations and the VDM 
  \cite{Hoehler76,Mergell96} were
  performed by several groups. These fits included much more theory input, and
  were constrained by the need to fit all four nucleon form factors.
  The most recent value resulting from such fits is the one of Mergell {\em et
  al.}, $0.847 \pm 0.009 fm$. The average, $0.854
  \pm 0.012~fm$, of this radius and the one of Simon  \et~ is quoted as the ''best''
   value in the compilation of Mohr and Taylor \cite{Mohr00}.
  
  Recent studies have provided additional insight: even for a system as light as the
  proton, Coulomb distortion of the electron waves needs to be accounted for
  \cite{Sick96b,Sick98}. This Coulomb distortion was shown to solve a long 
  standing puzzle with
  the deuteron \rms , and Rosenfelder demonstrated 
  \cite{Rosenfelder00} that it  also  increases
  the proton \rms . Using a restricted set of data and the above
  mentioned polynomial expansion he showed that the radius increases by about
  $0.01~fm$ when accounting for Coulomb distortion. 
  
  {\bf Model-independent radii?~~}
  In general, the groups studying the proton data have tried to extract 
  a \rms~ that is model-independent. This is possible when using
  as in eq.(1) the expansion of $G_e(q)$ in terms of the moments \rt, \rf,.. .
  At very low $q$, one could hope that the $q^4$\rf-term is small, such that
  the \rt - term can be determined without using a specific model for
  $G_e(q)$.
  
  This is true in principle, but very hard in practice. At small $q$ also
  the $q^2$\rt /6-term is small, and it is difficult to determine it accurately
  from the experimental form factors which are proportional to
  $1-$\qrt $+$... .
  Small systematic errors in the normalization of the cross sections 
    have a strong  influence on the small \qrt -term. When  ''eliminating''
  problems with the normalization of the data by floating them much of the
  sensitivity to the \rms~ gets lost and the norm-determining (implicit) 
  extrapolation to
  $q=0$  becomes very sensitive to small $q$-dependent systematic errors 
  in the data (which are always  ignored).
  
  In practice, one therefore has to include data at not-so-low $q$ which are 
  also sensitive to the higher moments. 
  The problem with theses moments is particularly detrimental for the proton. 
  The proton has approximately an exponential charge density (or, more
  accurately speaking, a form factor of the dipole shape, $G_e(q) = 
  (1+q^2 0.055fm^2)^{-2}$, the Fourier transform of which gives an exponential).
  For such a density (form factor) the higher moments are increasing with order,
  {\em i.e.} \rf= 2.5 \rt$^2$,  $\langle r^6 \rangle$ =11.6 \rt$^3$ etc, hence
  giving a large contribution to $G(q)$.
  
  The consequence: there is no $q$-region where the \rt~ term dominates the
  finite size effect to $>$98\%  {\em and}  the finite size effect is
  sufficiently big compared to experimental errors to allow a, say, 2\%
  determination of the \rms . There is also no region of $q$ where the \rf~
    moment can be determined accurately without getting into difficulty with the 
  $\langle r^6 \rangle$  term. Towards higher  $q$, the polynomial expansion is seriously
  restricted by the convergence radius of $\sim 1.4 fm^{-1}$.

\begin{figure}[hbt]
\begin{center}
\includegraphics[scale=0.5,clip]{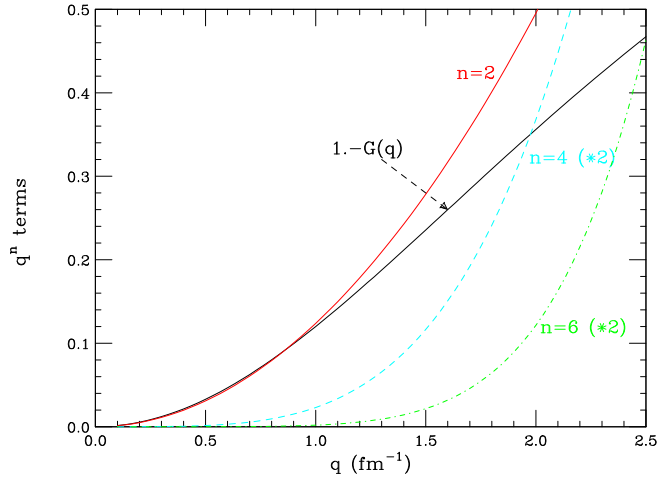}
\parbox{13cm}{\caption[]{
The figure shows the contribution of the $q^n$ terms to the finite size effect,
calculated using the moments from the CF parameterization.
The black curve  gives  the total finite-size effect. 
}\label{term}} 
\end{center} 
\end{figure}
  
  This situation is illustrated in fig.\ref{term} which shows the
  contribution of the various $q^n$ terms to the finite size effect. This
  problematic situation with the higher moments 
  is at the origin of the difficulties of determining 
  a model-independent proton \rms .     
          
{\bf Continued-fraction expansion.~~}
Continued Fraction (CF) expansions   
%
\begin{equation}
 G_{e}(q) = \frac{1 \hfill}{\displaystyle 1
                    + \frac{q^2b_1 \hfill }{\displaystyle 1
		       + \frac{q^2b_2 \hfill }{\displaystyle 1
		          + \cdots}}}
\end{equation}			
are a subclass of Pad\'e approximants which have initially been introduced to
solve the ''problem of moments'', {\em i.e.}  to find a function $f(z)$
specified by its moments $\langle z^n \rangle$ \cite{Jones80} and to accelerate 
the convergence of poorly converging series \cite{Haenggi80}. The radius of 
convergence of the CF expansion
is much larger than the one of the polynomial expansion, although within the
convergence radius of the latter it agrees exactly with it.

The moments of interest are directly linked to the coefficients $b_1, b_2, ..
b_N$ \
{\em i.e.} the coefficients of $q^2$, $q^4$,...  are given by   $b_1^2$, 
$b_1^2+b_1 b_2$,... . An important advantage, already exploited in fits of the 
deuteron form factor \cite{Klarsfeld86}, is the fact that the parameters 
$b_1$, $b_2$  for exponential-type densities are well decoupled. This is a
consequence of the fact that the CF is the natural parameterization for form
factors resulting from exchange-poles at $q^2 < 0$, the physical mechanism 
exploited in the VDM.

{\bf Tests of CF-expansion.~~}
In order to study the  dependence introduced by the usage of the CF
expansion with given number N of terms and given $q_{max}$, we have 
used pseudo-data. 
These cross sections were generated using  parameterized expressions
for the form factors (dipole form, or the dispersion relation parameterization 
of Hoehler {\em et al.} \cite{Hoehler76}). The pseudo data were generated at the
energies and angles of the experimental data, with the error bars of the 
experimental
data. In the fits, the pseudo data were used as calculated from the
parameterization, or with random fluctuations
calculated from the experimental error bars superimposed.  
 
 Fits of these pseudo-data were performed with the CF expansion with a variable
 number N of terms, and with variable  $q_{max}$ 
 of the points fitted. We have studied the scatter of the resulting fitted
 \rt ~ values, and their deviation from the known radius used in the
 generation of the pseudo-data. In these tests, we have been rather generous in
 accepting fits, {\em i.e.} by including fits with $\chi^2 \leq 1.2
 \chi^2_{min}$.
 
 When using the region $1 fm^{-1} < q_{max} < 5 fm^{-1}$ and 2 to 5 terms in
 the CF-expansion, we find a scatter of the fitted \rmss~ of $\pm 0.010 fm$
 around the true (input) values. This scatter we take as representative of the
 uncertainty due to the choice of N and $q_{max}$; it covers the statistical error (which for pseudo- and
 real data is the same by construction) as well.
 
 {\bf Analysis of world data.~~}
 In order to determine the proton \rms~ we use the world cross sections
 \cite{Bumiller61}-\cite{Bartel70} for $q<4 fm^{-1}$. The most precise data 
 relevant for the radius determination have been measured at Mainz 
   \cite{Borkowski74}-\cite{Simon81}. 
 These data are {\em absolute}, that is they
 have small systematic uncertainties in the absolute normalization. This type of
 data is the most useful one for a determination of the \rms . 
 
 We use for our fits the primary cross sections. When parameterizing both
 $G_e(q)$ and $G_m(q)$ with the CF expansion and fitting $G_e$ and $G_m$
 simultaneously to the cross sections, the L/T-separation is automatically
 performed, with superior quality as compared to the standard approach of
 separating L and T for each individual experiment.
 
 The Coulomb corrections are calculated in second-order Born approximation 
 according to \cite{Sick98} using an exponential charge  density.   
 These corrections are applied to the cross section  data, such that the 
 subsequent fit can be
 performed in PWIA as has been done in the past.
 
 In the fits we use all data with their standard random uncertainties. The error
 matrix is used to compute the random uncertainty of derived quantities. In
 order to evaluate the effect of the systematic uncertainties (normalization
 uncertainties) the individual data sets are changed by their quoted
 uncertainties, refitted and the resulting changes quadratically added.
 
 In the fits one finds experimental data sets (for instance the 40 years old 
 Stanford data) that have much too large a
 $\chi^2$; these points, however,  do not inappropriately influence the final
 result, so we have not increased their error bars just to get a good-looking 
 $\chi^2$.  We also find small discrepancies in the
 overall normalization of some data sets ({\em e.g.} the 
  data set of ref. \cite{Simon81} seems  $\sim$1\%
 high). We have chosen to keep the norm at the experimental value, and not float
 the data. For such precision experiments more than half the effort has gone 
 into
 the determination of the overall normalization; ignoring  this effort  by
 floating the norm (or greatly mitigating its influence by treating the
 normalization as just one further data point) does not do 
 justice to the experiments
 and leads to  loss of much information. Again, the effects upon the \rms~ 
 of the observed ''discrepancies'' have been found to be small and are 
 covered by the quoted uncertainty.
 
 As a check we have also used the polynomial expansion, with $q_{max} = 1.2
 fm^{-1}$ and the $q^4$ coefficient taken from a fit that explains the
 higher-$q$ data. We find the same \rms ~ as with the CF fit, but a larger 
 uncertainty and a  higher sensitivity to the $q_{max}$ employed. 
 
\begin{figure}[hbt]
\begin{center}
\includegraphics[scale=0.6,clip]{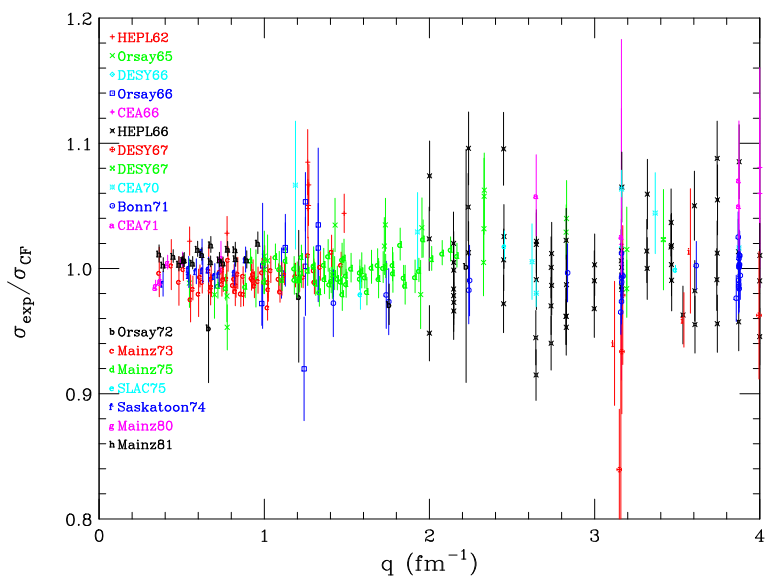}
\parbox{13cm}{\caption[]{
The figure shows the ratio of experimental and fit cross sections for the CF
parameterization. 
}\label{rat}} 
\end{center} 
\end{figure}

 \enlargethispage{5mm}
 The quality of the fits is quite good. We show in fig.\ref{rat} the ratio of
 experimental cross sections and fit for the CF parameterization and 5 CF
 coefficients. The $\chi^2$ is 512 for  310 data points \footnote{The $\chi^2$
 would reduce to 370 when adding quadratically 3\% to the Stanford error bars,
 with an increase of $r_{rms}$ of $0.002fm$. A norm change of 1\% of 
 \cite{Simon81}  would increase $r_{rms}$ by $0.007fm$ and decrease
  $\chi^2$ by 60.}. 
 The resulting \rms~ is $0.895 fm$. The uncertainty due to N, $q_{max}$ and
 statistics  
 is $\pm0.010 fm$, the systematic uncertainty $0.013fm$. This yields as the final
 result for the charge radius of the proton $r_{rms}^e = 0.895 \pm 0.018 fm$.
 This radius is significantly larger than the values generally cited 
 in the literature.
  It agrees with the most accurate value derived from atomic transitions
 \cite{Udem97} $0.890 \pm 0.014 fm$.

 {\bf Differences to previous determinations.~~}
 It may be interesting to understand why previous analyses gave smaller radii. 
 Simon {\em et al.} \cite{Simon80} ($r_{rms}=0.862fm$) used the polynomial expansion 
 up to $q^4$ and  $q_{max}=1.2fm^{-1}$, but
 found a \rf - moment that was a factor of ten smaller than given by fits
 that explain the proton data to higher $q$; this difference comes from very
 small systematic problems in the data which we have not further explored. 
 When repeating their fit with 
 the \rf - moment given by a fit that explains the data to larger $q$, 
 {\em e.g.} the one from the CF fit, one finds a
 radius that agrees with the one we find. 
 
 The fits based on dispersion relations and the VDM are strongly constrained by
 theory and the need to fit all four nucleon form factors. When looking at the
 ratio of experimental and VDM cross sections with the resolution 
 employed in fig.~\ref{rat} 
 the systematic deviations of the fits \cite{Hoehler76,Mergell96} 
 from the data at low $q$ are immediately obvious.
 
 Rosenfelder \cite{Rosenfelder00} ($r_{rms}=0.880fm$), whose primary interest was the 
 exploration of the effect
 of Coulomb distortion, also used the polynomial expansion, with the \rf~
 term taken from a low-$q$ fit quoted in the literature. When correcting his 
 value for a better
 \rf~ value from a good fit to the higher-$q$ data  
 and accounting for  differences in the data set, one arrives
 at the value of the proton \rms~ we find.
    
  {\bf Conclusions.~~} From an analysis of the {\em world}-data on e-p
  scattering we determine the proton \rms~ and find a value that is significantly
  larger than previous values. The change is understood as a consequence of
  treating properly the higher moments  $\langle r^n \rangle$.
     
 {\bf Acknowledgment.~~} The author acknowledges discussions with Savely
 Karshenboim which triggered this study.


\end{document}